\documentstyle[epsfig]{texas}

\begin{document}

\title{CLUMPING OF CDM FROM THE COSMOLOGICAL QCD TRANSITION}

\author{Dominik J.~Schwarz,$^1$ Christoph Schmid,$^2$ and Peter Widerin$^2$}
\smallskip

\address{(1) Institut f\"ur Theoretische Physik \\
             Robert-Mayer-Str.~10, Postfach 11 19 32,  
             D-60054 Frankfurt am Main, Germany \\
             {\rm Email: dschwarz@th.physik.uni-frankfurt.de} 
}\smallskip

\address{(2) Institut f\"ur Theoretische Physik \\
             ETH-H\"onggerberg, CH-8093 Z\"urich, Switzerland 
}

\begin{abstract}
The cosmological QCD transition affects primordial density perturbations.
If the QCD transition is first order, the sound speed vanishes during the
transition and density perturbations fall freely. For scales below the
Hubble radius at the transition the primordial Harrison-Zel'dovich spectrum
of density fluctuations develops large peaks and dips. These peaks grow with
wave number for both the radiation fluid and for cold dark matter (CDM). The 
peaks in the radiation fluid are wiped out during neutrino decoupling. For 
cold dark matter that is kinetically decoupled at the QCD transition 
(e.g.\ axions) these peaks lead to the formation of CDM clumps of masses 
$10^{-20} M_\odot< M < 10^{-10} M_\odot$.
\end{abstract}

\section{Introduction}

The transition from hot deconfined quarks and gluons to hot hadrons
happens in the early Universe at a temperature $T_\star \sim 150$ MeV 
\cite{MILC}. Lattice QCD results for the physical values of the quark 
masses indicate that the QCD phase transition is of first order \cite{Iwasaki}.

For a first-order QCD phase transition the sound speed vanishes during the 
coexistence of the deconfined and confined phases \cite{SSW}. The important 
scales are the mean distance between the hadronic bubbles at their nucleation, 
$R_{\rm nuc}$, which is a few cm for homogeneous nucleation \cite{CM,I}, 
and the Hubble scale $R_{\rm H} \sim m_{\rm P}/T^2_\star \sim 10^4$ m. 
Cosmological density perturbations are affected by the QCD transition at 
scales $\lambda \stackrel{<}{\scriptstyle\sim} R_{\rm H}$. 
At the coexistence temperature $T_\star$ the pressure $p_\star = p(T_\star)$
is fixed. Therefore pressure gradients and the sound speed vanish for
wavelengths much larger than the typical bubble separation. This gives rise 
to large peaks and dips above the primordial spectrum of density perturbations
in the radiation fluid. These peaks grow at most linearly with wave number. 

At $T\sim 1$ MeV the neutrinos decouple from the radiation fluid. During this 
decoupling the large peaks in the radiation spectrum are wiped 
out by collisional damping \cite{Weinberg,SSW2}.

Today the universe is dominated by dark matter, most likely cold dark
matter (CDM). If CDM is kinetically decoupled from the radiation fluid
at the QCD transition, the density perturbations in CDM do not suffer
from the neutrino damping. Examples for kinetically decoupled CDM are
primordial black holes, axions, or WIMPzillas, but not supersymmetric dark 
matter and heavy neutrinos \cite{SSW2}.
 
At the time of the QCD transition the energy density of CDM is small, i.e.\
$\rho^{\rm CDM}(T_\star) \sim 10^{-8} \rho^{\rm RAD}(T_\star)$. Kinetically
decoupled CDM falls into the potential wells provided by the dominant radiation
fluid. Thus, the CDM spectrum is amplified on subhorizon scales. The peaks 
in the CDM spectrum go nonlinear shortly after radiation-matter equality. 
This leads to the formation of CDM clumps with masses $10^{-20} M_\odot < M < 
10^{-10}M_\odot$ \cite{SSW2}.

If the QCD transition is strong enough, these clumps could be detected by 
gravitational femtolensing \cite{femtolensing}. The clumping of kinetically 
decoupled CDM implies that axions, if they are the CDM, could hide from 
axion search experiments \cite{Sikivie}. 

\section{Kinetically decoupled cold dark matter}

Let us discuss the properties of some popular CDM candidates at the QCD scale.

The lightest supersymmetric particle, most probably the {\em neutralino}, 
is one of the candidates for CDM \cite{Griest}. Constraints from LEP 2 and 
cosmology, together with the assumption of universality at the GUT scale, show 
that the neutralino mass is $m_\chi > 42$ GeV \cite{Ellis}. It is essential 
to distinguish between the chemical freeze-out and the kinetic decoupling of 
neutralinos. The chemical freeze-out happens when the annihilation rate of 
neutralinos drops below the Hubble rate, $\Gamma_{\rm ann}/H < 1$. At 
freeze-out the rate for neutralino annihilation, $\Gamma_{\rm ann} = 
\langle v\sigma_{\rm ann}\rangle n_\chi$, is suppressed by the Boltzmann 
factor in the number density of the neutralinos, $n_\chi \sim 
(m_\chi T)^{3/2}\exp(-m_\chi/T)$. The freeze-out temperature of
the neutralino is $T_f \sim m_\chi /20 > 2$ GeV.

Kinetic decoupling, in contrast, is determined by the elastic scattering
between neutralinos and the dominant radiation fluid. The interaction
rate for elastic scattering is $\Gamma_{\rm el} = \langle v\sigma_{\rm el} 
\rangle n$, where $n \sim T^3$ is the number density of relativistic particles.
The kinetic decoupling of the neutralino happens at the $10$ MeV scale 
\cite{SSW2} and therefore neutralinos belong to the radiation fluid at the 
QCD transition. The same holds true for {\em heavy neutrinos} with $m_Z /2 < 
m_\nu < 1$ TeV. 

Superheavy WIMPs, {\em WIMPzillas} with masses $10^{12}$ -- $10^{16}$ GeV 
\cite{Chung}, decouple from the radiation fluid well before the QCD transition.
Although WIMPzillas scatter with the radiation fluid at $T_\star$, the momentum 
transfer in these interactions is tiny compared to the WIMPzilla's momentum. 
Even the large number of scatterings per Hubble time cannot change the 
momenta of WIMPzillas significantly (if $m_{\rm WIMP} > 10^6$ GeV). Thus, 
WIMPzillas cannot be dragged along by the radiation fluid.
 
{\em Axions} could be the dominant matter today if their mass is small, i.e.\ 
$m_{\rm a} \sim 10^{-5}$ eV, which corresponds to a breaking of the 
Peccei-Quinn (PQ) symmetry at the scale $f_{\rm PQ} \sim 10^{12}$ GeV
\cite{rev}. These axions could be produced coherently due to a initial
misalignment of the axion field and by the decay of axionic strings.
The initially misaligned axion field starts to oscillate coherently when
the axion mass has grown to $m_{\rm a}(T_1) \sim 3 H(T_1)$, where $T_1
\sim 1$ GeV. Below $T_1 \sim 1$ GeV the oscillating axion field evolves as CDM.

If the reheating temperature after inflation lies above $f_{\rm PQ}$, 
the axion field is inhomogeneous on scales larger $R_{\rm H}(T_1)$.
These inhomogeneities produce axion miniclusters with mass $M_{\rm mc} 
\sim 10^{-12} M_\odot$ and radius $R_{\rm mc} \sim 0.1 R_\odot$ today 
\cite{Hogan}.

A further candidate for CDM that is kinetically decoupled at the
QCD scale is {\em primordial black holes} (PBH) \cite{CarrHawking}
produced before the QCD transition and therefore with masses $M_{\rm BH}
\ll 1 M_\odot$. In order to survive until today PBH should have $M_{\rm BH}
> 10^{15}$g $\approx 10^{-18} M_\odot$. PBH in the range from $10^{-18}
M_\odot$ to $10^{-16} M_\odot$ would radiate too strongly to be compatible
with $\gamma$-ray observations \cite{Carr}. 

\section{Sound speed}

The QCD phase transition starts with a short period ($10^{-4} t_{\rm H}$)
of tiny supercooling, $1-T_{\rm sc}/T_\star\sim 10^{-3}$. When $T$ reaches 
$T_{\rm sc}$, bubbles nucleate at mean distances $R_{\rm nuc} 
\stackrel{<}{\scriptstyle\sim} 2$ cm \cite{CM}. The bubbles grow most 
probably by weak deflagration \cite{I}, because the surface tension
is very small \cite{ls}. The released energy is transported into the 
deconfined phase by shock waves, which reheat the deconfined phase to 
$T_\star$ within $10^{-6} t_{\rm H}$. The entropy production during this 
very short period is $\Delta S/S \sim 10^{-6}$. Further bubble nucleation 
in the reheated quark-gluon phase is prohibited after this first $10^{-4} 
t_{\rm H}$. Thereafter, the bubbles grow adiabatically due to the expansion 
of the Universe (coexistence of the phases). The transition completes after 
$10^{-1} t_{\rm H}$.

During the reversible coexistence period (isentropic) $T$ and $p$ are fixed,
while $\rho$ can vary, therefore
\begin{equation}
\label{cs2}
c_s^2 \equiv \left(\partial p\over \partial
\rho\right)_{\rm isentropic} = 0 \ .
\end{equation}
This implies that there are no pressure gradients at scales $\lambda >
R_{\rm nuc}$, while both phases coexist during the first-order phase
transition.

The thermodynamics of the cosmological QCD transition can be
studied in lattice QCD, because the baryochemical potential is
negligible in the early Universe, i.e.\ $\mu_{\rm b}/T_\star \sim 10^{-8}$.
We fit the equation of state from lattice QCD \cite{Karsch}.

\section{The evolution of density perturbations}

During inflation density perturbations
$\delta\!\rho$ have been generated with an almost scale-invariant 
Harrison-Zel'dovich spectrum \cite{perturbations}. In the radiation 
dominated era subhorizon ($k^{\rm phys} \gg H$) density perturbations 
oscillate as acoustic waves, supported by the pressure 
$\delta p = c_s^2 \delta\!\rho$. It is useful to 
study the dimensionless density contrast $\delta \equiv \delta\!\rho/\rho$
as a function of conformal time and comoving wave number. The subhorizon
equation of motion reads
\begin{equation}
\label{delta}
\delta'' + c_s^2 k^2 \delta = 0 \ ,
\end{equation}
if we assume that the phase transition is short compared to the Hubble 
time \cite{SSW2}.
From the lattice QCD equation of state the period of vanishing
$c_s$ lasts for $0.1 t_{\rm H}$, in the bag model it lasts
for $0.3 t_{\rm H}$.

To illustrate the mechanism of generating large peaks above the HZ spectrum
we discuss the subhorizon evolution of the density contrast $\delta$ in the
bag model.
Before and after the QCD transition $c_s^2 = 1/3$ and the density contrast
oscillates with constant amplitudes $A^{\rm in}$ resp.~$A^{\rm out}$.
At the coexistence temperature $T_\star$ the pressure gradient, i.e.\
the restoring force in Eq.~(\ref{delta}), vanishes and
the sound speed (\ref{cs2}) drops to zero.
Thus $\delta$ will grow or decrease linearly, depending on the fluid velocity
at the moment when the phase transition starts. This produces large peaks
in the spectrum which grow linearly with the wave number, i.e.\
$A^{\rm out}/A^{\rm in}|_{\rm peaks} = k/k_1$.

Kinetically decoupled CDM falls into the gravitational potential wells 
of the radiation fluid during the coexistence regime.

\begin{figure}
\centering
\epsfig{figure=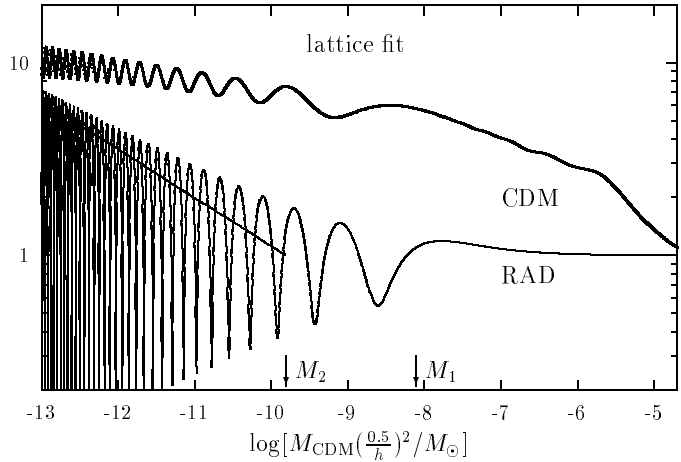, width=9cm}
\caption{The transfer functions for the density contrast of CDM and of 
the radiation fluid due to the QCD transition. On the horizontal axis 
the wave number $k$ is represented by the CDM mass contained
in a sphere of radius $\pi/k$. } \label{fig1}
\end{figure}

Fig.~\ref{fig1} shows the final spectra for the density contrast
of the radiation fluid and of CDM for a fit \cite{SSW} to lattice QCD.
The spectra have been obtained from numerical integration of the fully 
general relativistic equations of motion. The mass $M_1$ corresponds to
the wave number $k_1$ of the bag model analysis. $M_1$ coincides with 
the CDM mass inside the Hubble horizon at the QCD transition, $\sim 10^{-8} 
M_\odot$. For the equation of state from the fit to lattice QCD a WKB 
analysis shows that $A^{\rm out}/A^{\rm in}|_{\rm peaks} = (k/k_2)^{3/4}$,
which defines the mass $M_2$. 
We conclude that at the horizon scale the modification of the HZ
spectrum is mild, whereas for scales much smaller than the horizon
big amplifications are predicted.  Without tilt in the COBE normalized
spectrum the density contrast grows nonlinear for 
$k^{\rm phys}/H \stackrel{>}{\scriptstyle \sim}
10^4$ resp.~$10^6$ for the bag model resp.~lattice QCD equation of state.

\section{Clumps in CDM}

At $T\sim 1$ MeV the neutrinos decouple from the Hubble scale. During their 
decoupling they wash out all fluctuations of the radiation fluid below 
a CDM mass of $10^{-5} M_\odot$. Thus, at the time of nucleosynthesis 
all peaks in the spectrum of the radiation fluid (Fig.~\ref{fig1}) are erased
and the energy density on scales $\lambda \stackrel{<}{\scriptstyle \sim}
R_{\rm H}(1$ MeV$)$ is homogeneous.

For kinetically decoupled CDM collisional damping is irrelevant. 
Peaks in these types of CDM survive and grow logarithmically during the 
radiation era. Shortly after equality they grow nonlinear and collapse by 
gravitational virialization to clumps of $M < 10^{-10} M_\odot$. From
a COBE normalized HZ spectrum and the equation of state from quenched
lattice QCD a clump of mass $10^{-15} M_\odot$ would have a
size of $\sim 10 R_\odot$. For a tilted spectrum with $n = 1.2$ the size of the
same clump would be $\sim 1 R_\odot$ \cite{SSW2}. 

We find that the peaks in the CDM spectrum lead to clumps
of masses $10^{-20} - 10^{-10} M_\odot$. Today, these clumps would
have a density contrast of $10^{10} - 10^{17}$, where the lower value
corresponds to a $10^{-15} M_\odot$ clump from an untilted CDM spectrum, the
bigger value is for a $10^{-20} M_\odot$ clump from a tilted CDM spectrum.
The evolution of these clumps in the late stages of structure formation
remains to be investigated (disruption, mergers, etc.).

For a larger amplification during the QCD transition, e.g., if it should 
turn out that the latent heat is bigger than the value from present 
lattice QCD calculations, more compact clumps are possible. These could be 
subject to femto-lensing \cite{femtolensing}.
With the values from lattice QCD, the CDM clumps are not
compact enough to lie within the Einstein radius, which is $R_{\rm E}
\sim 0.02 R_\odot$ for a $10^{-15} M_\odot$ clump.

The clumping of CDM changes the expected rates for some dark matter 
searches, because some of the rates depend on the space-time position of 
the detector, star, or planet. Especially experiments looking for axion 
decay in strong magnetic fields \cite{Sikivie} would not provide a new limit on 
the axion mass if they find nothing. These experiments may just tell us 
that we are not sitting in an axion clump currently. 
These consequences remain to be studied further.

\section*{Acknowledgments}
D.J.S thanks the Alexander von Humboldt foundation for financial support.

\end{document}